\newcommand\lsim{\lesssim}

\documentclass{emulateapj}
\usepackage{apjfonts}

\begin{document}

\title{The structure of dark matter halos: self-similar models versus N-body simulations}
\author{Leonid Chuzhoy}
\bigskip
\affil{McDonald Observatory and Department of Astronomy, The University of Texas at Austin, RLM 16.206, Austin, TX 78712, USA; 
chuzhoy@astro.as.utexas.edu}

\begin{abstract}
We derive the density profile of cold dark matter halos using a self-similar accretion model. 
We show that if the clumpiness of the infalling matter is taken into account, then the inner density slope, 
$\delta =d\log{\rho}/d\log{r}$, is close to $-1$. Compared with the density profiles predicted by different numerical simulations, 
we find that outside $\sim 0.1\%$ of the virial radius, our solutions agree best with the fitting formula proposed by Navarro et al. (2004),
$d\ln{\rho}/d\ln{r}=-2(r/r_{-2})^\alpha$, $\alpha\sim 0.17$, where $r_{-2}$ is a characteristic radius, inside which the density profile becomes shallower than isothermal ($\delta<-2)$.
\end{abstract}

\keywords{cosmology: theory -- dark matter -- large-scale structure of Universe}

\section{\label{Int}Introduction}
In the standard $\Lambda$CDM paradigm dark matter halos evolve out of small fluctuations in the initial density field.  Since the 
structure of virialized objects may depend on their formation history, many attempts have been made to link the observed density 
distribution of dark matter halos with the initial spectrum of density fluctuation.
High resolution N-body simulations indicate that density profiles of dark matter halos have a universal shape
\begin{eqnarray}
\label{NFW}
\rho(r)=\frac{\rho_s}{(cr/r_{vir})(1+cr/r_{vir})^2}
\end{eqnarray}
whose dependence on the initial conditions is apparent only from the value of the concentration parameter $c$ 
(Navarro, Frenk \& White 1997; henceforth NFW). While the NFW model has been usually successful in reproducing the observed mass 
profiles of large galaxies and galaxy clusters, it seems to be inconsistent with the rotation curves of the dwarf galaxies. In 
addition, several other N-body simulations \cite{Mo,FM,Ri}, while generally obtaining similar results at large radii, found different density slopes in the inner regions of halos, thus raising some doubts about the accuracy of the NFW model.

The alternative approach to N-body simulations is provided by analytic modeling and in particular by self-similar accretion models \cite{FG,Bert}. The choice of a particular set of initial conditions (the profile of an initial density perturbation is assumed to be a power law, $\Delta M/M\propto r^{-3\epsilon}$) allows to recover an exact analytic solution for the final density distribution, that can be compared to the N-body results. Further, it can be shown that the appropriate value of $\epsilon$ (and therefore the final density profile) is fixed by the spectrum of the initial density fluctuations, $\epsilon=(n+3)/6$, where $n(k)=d\log P(k)/dk$ \cite{P}. Unfortunately, the very simplicity that is a strong point of the analytic models, can also lead to their downfall, if an important physical process has been left out.  Purely radial infall that results from the evolution of a spherically symmetric density perturbation, produces an extremely steep inner density profile, $\delta=d\log \rho(r)/d\log r\leq -2$. As neither the observations no any of the  N-body simulations produced a similarly steep density distribution, attempts have been made to fix 
the self-similar models by giving collapsing halos some initial angular momentum \cite{WZ,STW,Sub,Nus}. Including angular momentum 
allows to reduce the steepness of the inner slope, down to $-9\epsilon/(3\epsilon +1)=-(3n+9)/(n+5)$.  However, in general the solutions, 
being dependent on the amount of added angular momentum,
still did not reproduce the universal shape given by equation (\ref{NFW}).

In this paper we show that the discrepancy between the self-similar models and N-body simulations is naturally resolved by including 
the process of dynamical relaxation into the former. We prove that even tiny deviations from spherical symmetry are sufficient to make 
the dark matter particles velocity distribution isotropic in the inner region, provided the density slope is steeper than $-1$. 
Furthermore, we show that for $n<-1$ ($\epsilon<1/6$) isotropic velocity distribution leads to the inner density slope $\delta\sim -1$. Since the 
spectral index, $n(k)$, spans the range from $-3$ (for smallest halos) to $-1$ (for galaxy clusters), our results imply that if the 
dark matter gravity is the only force to be considered,  then $\delta\sim -1$ in all existing virialized halos.

The paper is organized as follows. In \S 2, we show that density profiles, which are steeper than $\rho\propto r^{-1}$, are inconsistent with unisotropic velocities at the central region. In \S 3, we derive a density profile for a self-similar accretion model that includes dynamical relaxation process. In \S 4, we compare our results with the N-body codes. We summarize our conclusions in \S 5.

\section{Dynamical relaxation}  

In the standard $\Lambda$CDM paradigm the structure formation is hierarchical. The smallest halos that form first subsequently merge into larger and larger objects. If dynamical friction  between merging dark matter clumps is weak, clumpiness and radial orbits can be retained for a long time. On the other hand, if dynamical friction is strong, the clumps are destroyed and the velocities of dark matter particles become isotropic. As we show below, the latter is always the case at the central regions of halos with sufficiently steep ($\delta>-1$) density profiles.

Consider a halo of mass $M_h$ and radius $R_h$ formed by smaller clumps with the typical mass $m_{cl}$. The clumps characteristic relaxation time is
\begin{eqnarray}
\label{trel}
t_{rel}\sim &&(\pi n_{cl}G^2 m_{cl}^2V_{cl}^{-3}\log[M_h/m_{cl}])^{-1},
\end{eqnarray}
 where $n_{cl}\sim \rho/m_{cl}$ and $V_{cl}$ are the clump number density and characteristic velocity. The kinetic energy of clumps must be less than their potential energy at the center, $V_{cl}^2/2 \lsim GM_h/R_h$. This sets an upper limit on the relaxation time
\begin{eqnarray}
\label{trel2}
t_{rel}\lsim \frac{\sqrt{8 M_h^{3}}}{\pi n_{cl}G^{1/2} m_{cl}^2 R_h^{3/2} \log[M_h/m_{cl}]}.
\end{eqnarray}
Since the radial infall velocity of a clump is less than $\sim \sqrt{2GM_h/R_h}$, the time required for a clump to reach the center from a distance $r$, $t_{cr}$, exceeds $ r/ \sqrt{2GM_h/R_h}$.
For radial orbits to be sustained, the crossing time must be less than the relaxation time
\begin{eqnarray}
\label{trelcr}
1<\frac{t_{rel}}{t_{cr}}\lsim \left(\frac{4M_h^2}{\pi m_{cl} \log[M_h/m_{cl}]R_h^2}\right)[\rho r]^{-1}.
\end{eqnarray}
Obviously, if the density profile is steeper than $\rho\propto r^{-1}$,  then close to the center (i.e. for small values of $r$) the above inequality would be violated, implying that in such a case dynamical friction must be strong and radial orbits cannot be sustained. 
Therefore, {\it in the case where purely radial infall leads to a density slope below $-1$, dynamical relaxation must either isotropize the velocities or flatten the central density cusp to $\rho\propto r^{-1}$.}

\section{Self-similar infall}
We now proceed to calculating the actual density distribution of the collapsing halos, including the dynamical relaxation process. 
In the Einstein-de Sitter model (which is an excellent fit at high redshifts) a system evolving from a power law density 
perturbation ($\Delta M/M\propto r^{-3\epsilon}$) over a time approaches self-similarity, which allows to use dimensionless variables 
for mass profile, density, bulk velocity and energy density of dark matter particles.
\begin{eqnarray}
m(r,t)&=&M(\lambda)\frac{2r_{vir}^3}{9 G t^2}, \\
\rho(r,t)&=&D(\lambda)\frac{1}{6\pi G t^2}, \\
v(r,t)&=&V(\lambda)\frac{r_{vir}}{t}, \\
p(r,t)&=&\Pi(\lambda)\frac{r_{vir}^2}{18\pi G t^4},
\end{eqnarray}
where $\lambda=r/r_{vir}$ and $t$ is a Hubble time. The virial radius, $r_{vir}$, which is customarily defined as a radius inside which the density contrast is $18\pi^2\approx 178$, grows as $t^\eta$, where $\eta=2(3\epsilon+1)/9\epsilon$.

At a fixed $\lambda$ the relaxation time (equation (\ref{trel})) scales as $t$, which implies that the relaxation process does not break self-similarity.
However, for the sake of simplicity, rather than evaluate the impact of dynamical relaxation on particles velocities at each point, we assume 
that at large radii the infall is purely radial, neglecting small tangential motions.  Close to the center we assume that the velocities are completely isotropic. 
Furthermore, we assume that an abrupt transition between the two regimes occurs at some point.
This transition is analogous to the shockheating of the infalling gas (though for collisionless dark matter the width of the transition 
region can be significantly larger than the width of the shock, which roughly equals baryon mean free path). This approach allows us to 
treat the dark matter as a dissipationless fluid whose dynamics is determined by continuity, Euler and adiabatic equations. Cast into the 
dimensionless variables, these equations become
\begin{eqnarray}
\label{cont}
&&(V-\eta\lambda)D'+\left(\frac{2V}{\lambda}+V'-2\right)D=0, \\
&&(\eta-1)V+(V-\eta \lambda)V'=-\frac{\Pi'}{D}-\frac{2M}{9\lambda^2}, \\
&&\left(\frac{5D'}{3D}-\frac{\Pi'}{\Pi}\right)(V-\eta\lambda)=2\eta-\frac{2}{3}, \\
\label{mass}
&&M'=3\lambda^2D,
\end{eqnarray}
If all non-adiabatic cooling and heating processes can be neglected, the above equations are equally applicable to the baryonic and dark matter fluids \cite{CN}. 

Outside the shock the radial motion of the infalling matter can be traced analytically \cite{Bert}.
The jump conditions at the adiabatic shock are
\begin{eqnarray}
V^+&=&\eta\lambda_s+\frac{V-\eta\lambda_s}{4}, \\
D^+&=&4D^-, \\
P^+&=&\frac{3}{4}D^-(V-\eta\lambda_s)^2 \\
M^+&=&M^-,
\end{eqnarray}
where the superscript minus and plus signs refer to pre- and post- shock quantities.
The location of the shock, $\lambda_s$, which is initially unknown, have to be found by integrating equations (\ref{cont})-(\ref{mass}) 
inwards. Too small values of $\lambda_s$ produce a non-zero mass at $\lambda=0$, while too large values of $\lambda_s$ give a singularity at some $\lambda>0$. A unique value of $\lambda_s$ can therefore be recovered by requiring the mass and the infall velocity to be zero at $\lambda=0$. Typically, we find that the value of $\lambda_s$ is close to the location of the first dark matter shell-crossing, which is obtained when dynamical relaxation is neglected.

The asymptotic analysis of equations (\ref{cont})-(\ref{mass}) reveals the existence of two different types of solutions. For $-3<n<-2$ ($\epsilon<1/6$) the matter energy density is finite everywhere and close to the center the density slope goes to $\delta=-3(n+7)/(n+17)$, thus spanning the range between $-6/7\approx -0.86$ and $-1$. Since completely isotropic velocities, which we assumed at the center, require  $\delta<-1$, it is unclear whether $\delta>-1$ can actually be obtained. However, the convergence of $\delta$ to its asymptotic limit at the center is rather slow (at $r=0.01r_{vir}$ the slope is still $\sim -1.3$ and $-1.2$ for $n=-2$ and $-2.5$, respectively), $\delta+3(n+7)/(n+17)\propto 1/\log{r}$. Thus, except for extremely small values of $r$, our solutions should remain valid.

For $n>-2$ the matter energy density goes to infinity at the center and the density slope goes to $-(3n+9)/(n+5)$.\footnote{Accidentally, when the dynamical relaxation process is neglected, the same slope is also obtained for $n>1$.} However, since $n>-2$ is obtained only on scales, whose evolution is still in the linear regime, these solutions do not apply to any existing halos.

\begin{figure}[t]
\resizebox{\columnwidth}{!}
{\includegraphics{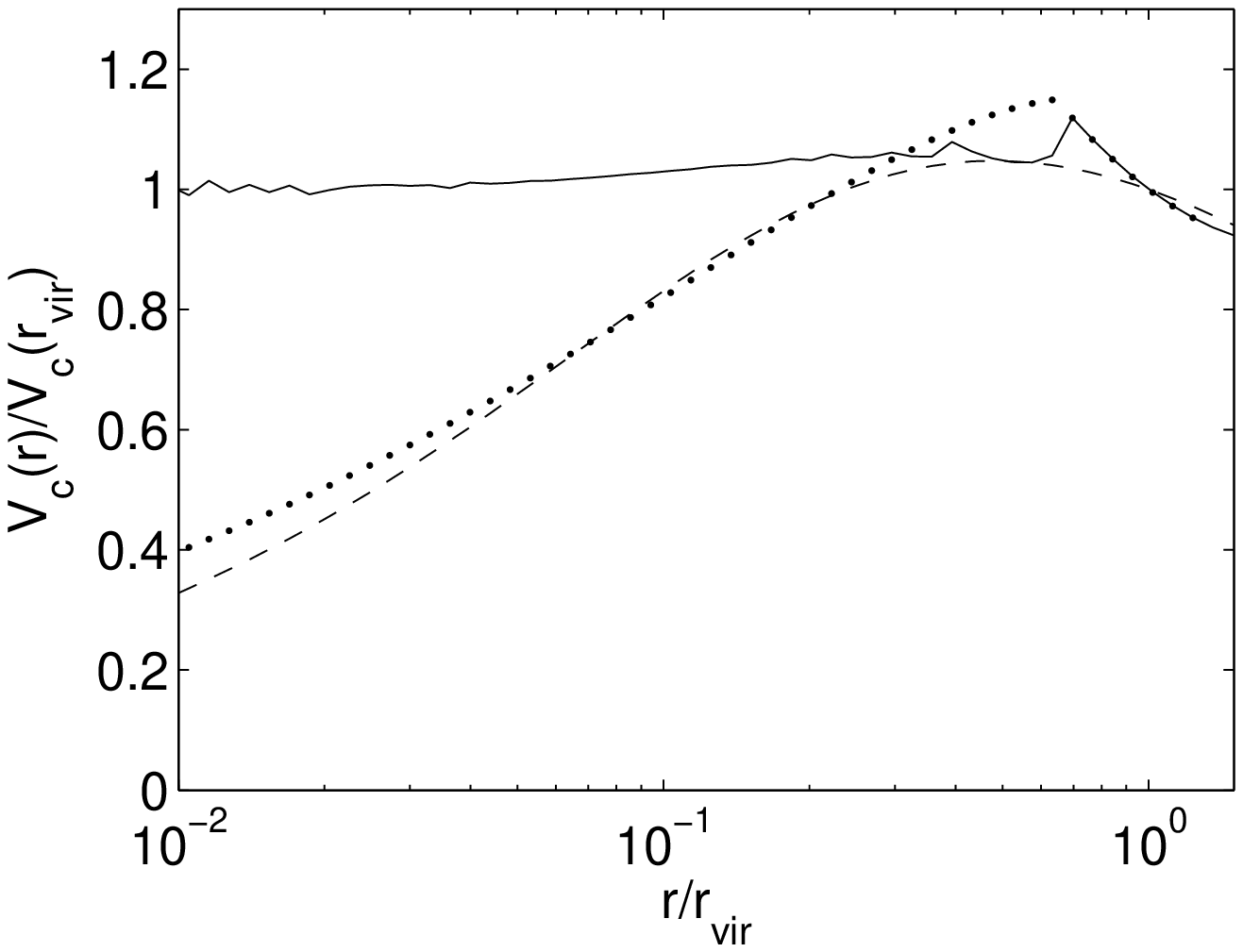}}
\caption{\label{fig1}Circular velocity $[V_c=\sqrt{Gm(r)/r}]$ profile for $n=-2$ ($\epsilon=1/6$). The dotted and the solild lines show the curves, respectively, with and without dynamical relaxation included. The dashed line shows the NFW profile for $c=4.5$.}
\end{figure}

\begin{figure}[t]
\resizebox{\columnwidth}{!}
{\includegraphics{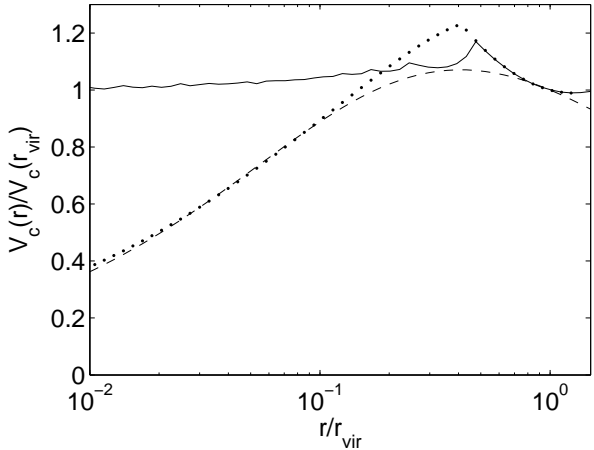}}
\caption{\label{fig2}Circular velocity $[V_c=\sqrt{Gm(r)/r}]$ profile for $n=-2.5$ ($\epsilon=1/12$). The dotted and the solild lines show the curves, respectively, with and without dynamical relaxation included. The dashed line shows the NFW profile for $c=5.3$.}
\end{figure}

\section{Self-similar solutions vs N-body codes}
\begin{figure}[t]
\resizebox{\columnwidth}{!}
{\includegraphics{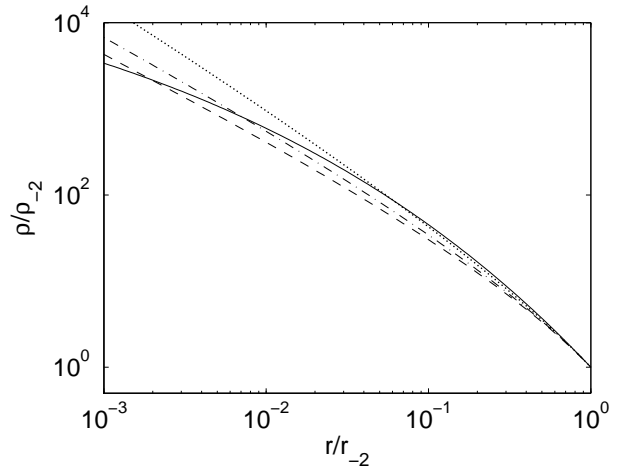}}
\caption{\label{fig3} The dark matter density versus radius. The dotted,  dashed-dotted and dashed lines correspond, respectively, to $n=-2$, $-2.5$ and $-2.75$. The solid line corresponds to the fitting function (Eq. \ref{pow}) with $\alpha=0.17$.}
\end{figure}

We find that for $0.01\lsim r/r_{vir}\lsim 1$ our solutions can be well approximated by the NFW fit (Figures \ref{fig1} and  
\ref{fig2}). The discrepancy of order $10\%$ in the circular velocity profile, which is seen at the outer part, where ``shock'' takes 
place, is quite expected, given our simplistic modeling of this transitional region. 
Closer to the center ( $0.0001 \lsim r/r_{vir}\lsim 0.01$), our solutions predict steeper density slope than the NFW fit (though not as 
steep as predicted by Moore et al. (1999)).
This is consistent with the later simulation results of Navarro et al. (2004), who suggested an improved fitting formula for the density profile
\begin{eqnarray}
\label{pow}
\ln\left(\frac{\rho(r)}{\rho_{-2}}\right)=-\frac{2}{\alpha}\left[\left(\frac{r}{r_{-2}}\right)^\alpha-1\right],
\end{eqnarray}
where  $r_{-2}$ is the radius inside which the density slope is above $-2$ and  $\rho_{-2}=\rho(r_{-2})$. We find that at small radii the equation (\ref{pow}) is a better approximation to our results (see Figure \ref{fig3}) and the values of $\alpha$ that produce the best fits to our solutions in the range $0.001<r/r_{-2}<1$ ($0.16$, $0.21$ and $0.22$, respectively, for  $n=-2$, $-2.5$ and $-2.75$) are close to $\alpha=0.17$ found by Navarro et al. (2004). However, at  $r\lsim 0.001r_{-2}$ (which has not been resolved by the N-body codes) the agreement between the equation (\ref{pow}) and our results breaks down.

\section{Summary}
We have shown that the results of high-resolution N-body simulations of dark matter infall can be quite accurately reproduced by a 
simple self-similar accretion model. Moreover, the self-similar accretion model, being unlimited by computational constraints, may 
have an advantage over the N-body codes in reconstructing the density profile in the central region of dark matter halos. Naturally, 
it should be kept in mind that close to the center of a halo, a stellar disk or a supermassive black hole may strongly affect the 
density profile, making inadequate any model or code that rely exclusively on the dark matter gravity.

It is interesting to note that the two threshold values of $n$, which separate different types of self-similar solutions, have a special 
significance in the real Universe. Thus $n=-2$, which separates the solutions with finite and infinite central energy density, is also the largest possible value of $n$ on scales of existing virialized halos. The $n=1$ threshold, which separates the solutions with finite and infinite gravitational potential in the center, seems also to be the largest value of $n$ on all scales \cite{Sp}.

\acknowledgments
LC thanks the McDonald Observatory for the W.J. McDonald Fellowship.

\end{document}